\documentclass[traditabstract]{aa}  
\usepackage{natbib}
\bibpunct{(}{)}{;}{a}{}{,} 
\usepackage{graphicx}
\usepackage{txfonts}
\def\ptrs{Phil. Trans. R. Soc. London, Ser. A}

\newcommand{\va}{c_{\mathrm{A}}}
\newcommand{\vap}{c_{\mathrm{Ap}}}

\newcommand{\der}{{\rm d}}

\newcommand{\vk}{c_{\rm k}}
\newcommand{\vkp}{c_{\rm kp}}
\newcommand{\vke}{c_{\rm ke}}

\newcommand{\rp}{\rho_{\rm p}}
\newcommand{\re}{\rho_{\rm e}}
\newcommand{\rc}{\rho_{\rm c}}
\newcommand{\lp}{L_{\rm p}}

\begin{document}

	\title{Damped kink oscillations of flowing prominence threads}

	\titlerunning{Damped oscillations of flowing threads}

   \author{R. Soler\inst{\ref{leuven}} \and M.~S. Ruderman\inst{\ref{sheffield}} \and M. Goossens\inst{\ref{leuven}}}
\offprints{R. Soler}
\institute{Centre for Mathematical Plasma Astrophysics, Department of Mathematics, KU Leuven,
              Celestijnenlaan 200B, 3001 Leuven, Belgium \\
                       \email{roberto.soler@wis.kuleuven.be}
 \label{leuven}
 \and
 Solar Physics and Space Plasma Research Centre (SP$^2$RC), University of Sheffield, Hicks Building, Hounsfield Road, Sheffield S3 7RH, UK \label{sheffield} }

 	 \date{Received XXX / Accepted XXX}

\abstract
{Transverse oscillations of thin threads in solar prominences are frequently reported in high-resolution observations. Two typical features of the observations are that the oscillations are  damped in time and that simultaneous mass flows along the threads are detected. Flows cause the dense threads to move along the prominence magnetic structure while the threads are oscillating. The oscillations have been interpreted in terms of standing  magnetohydrodynamic (MHD) kink waves of the magnetic flux tubes which support the threads. The damping is most likely due to resonant absorption caused by plasma inhomogeneity.  The technique of seismology uses the observations combined with MHD wave theory to estimate prominence physical parameters. This paper presents a theoretical study of the joint effect of flow and resonant absorption on the amplitude of standing kink waves in prominence threads. We find that flow and resonant absorption can either be competing effects on the amplitude or both can contribute to damp the oscillations depending on the instantaneous position of the thread within the prominence magnetic structure. The amplitude profile deviates from the classic exponential profile of resonantly damped kink waves in static flux tubes. Flow also introduces a progressive shift of the oscillation period compared to the static case, although this effect is in general of minor importance. We test the robustness of seismological estimates by using synthetic data aiming to mimic real observations. The effect of the thread flow can significantly affect the estimation of the transverse inhomogeneity length scale. The presence of random background noise adds uncertainty to this estimation. Caution needs to be paid to the seismological estimates that do not take the influence of flow into account.}

     \keywords{Sun: filaments, prominences ---
		Sun: oscillations ---
                Sun: corona ---
		Magnetohydrodynamics (MHD) ---
		Waves}

   \maketitle


\section{Introduction}

Solar prominences and filaments are large-scale magnetic structures of the solar corona \citep[see the recent reviews by][about the physics, dynamics, and modelling of prominences]{labrossereview,mackay}. High-resolution H$\alpha$ observations reveal that prominences are formed by a myriad of fine structures usually called threads \citep[see, e.g.,][]{lin2011}. Threads are thin and long plasma condensations which outline the prominence magnetic field. Theoretically, prominence threads have been modelled as magnetic flux tubes anchored in the solar photosphere \citep[e.g.,][]{ballesterpriest,rempel}, which are only partially filled with the cool ($\sim 10^4$~K) prominence material, i.e., the thread itself, while the rest of the tube is occupied by hot coronal plasma and therefore invisible in H$\alpha$ images. 

 Threads are highly dynamic. For example, transverse thread oscillations and propagating waves along the threads are frequently observed in prominences, which have been interpreted in terms of magnetohydrodynamic (MHD) kink waves \citep[see the reviews by][]{arreguiballester,arreguilivingrev}. The reported periods are usually in between 2 and 10 minutes, and the oscillations are typically damped after a few periods \citep{ning}. Resonant absorption, caused by plasma inhomogeneity in the transverse direction to the magnetic field has been proposed as the damping mechanism \citep[e.g.,][]{arregui08,arregui2d,solerslow,solerstatic}. In addition, flows and mass motions in prominences have been also reported \citep[e.g.,][]{zirker98, lin2003,okamoto}. The mean flow velocities are in most cases less than 30~km~s$^{-1}$, although values up to 40--50~km~s$^{-1}$ have been observed in active region prominences. The presence of flow can have a direct impact on the behavior of the waves \citep[see a discussion on this issue in][]{carbonell2009}. The combined use of both observations and MHD wave theory allows to apply the technique of seismology to solar prominences \citep[see][]{oliver,arregui12a}.  

Some observations point out the simultaneous presence of transverse oscillations and mass flows in prominence threads. The work by \citet{okamoto} is probably the best example.  \citet{okamoto} observed an active region prominence with Hinode/SOT using Ca II H-line images. They detected that some threads in the prominence were flowing presumably along the magnetic field  with an apparent velocity on the plane of sky of around 40~km~s$^{-1}$. Simultaneously, the threads were oscillating in the transverse direction. The mean period of the oscillations was 3~min. The oscillations were in phase along the whole length of the threads, which was roughly between 3,000~km and 16,000~km. Thus, a standing wave interpretation was proposed and the actual wavelength was estimated to be at least 250,000~km, which would correspond to twice the total length of magnetic field lines, approximately.

The observations by \citet{okamoto}  motivated a number of theoretical works that aim to interpret the observed thread dynamics.  \citet{terradas2008}  interpreted the observations by \citet{okamoto} in terms of standing MHD kink modes and used the observed periods to perform a seismological estimation of a lower bound of the prominence Alfv\'en speed. Subsequently, \citet{solerthreadflow} revisited the same event and performed both an analytical and a numerical investigation of the influence of flow on the period and amplitude of the standing waves. However, neither \citet{terradas2008} nor  \citet{solerthreadflow} took the damping of the oscillations into account. \citet{okamoto} did not study the evolution in time of the amplitude of the transverse thread oscillations, but damping has been reported in other events  \citep[see, e.g.,][]{ning}. The main purpose of the present work is therefore to determine the joint effect of flow and resonant absorption on the amplitude of standing kink MHD modes in prominence threads. The resonant damping of kink waves in the presence of flow has been studied in the past in the context of coronal loops \citep[see, e.g.,][]{goossens92,terradas2010,soler2011}, but not in the case of prominences.

We apply the general analytic theory developed by \citet{ruderman2011a,ruderman2011b} for standing kink waves in magnetic flux tubes with a time-dependent background. Analytic expressions for the kink mode amplitude and period as functions of the model parameters are derived. In addition, since the efficiency of resonant damping is directly controlled by the transverse inhomogeneity length scale of the threads, part of the present paper is also devoted to test the impact of flow on the estimation of this parameter using the technique of seismology \citep{goossens08}.

This paper is organized as follows. Section~\ref{sec:model} contains a description of the prominence thread model used in this work. The mathematical method is presented in Section~\ref{sec:maths}, where the main equations are obtained. Then, the impact of the different model parameters on the amplitude of kink modes is studied in Section~\ref{sec:param}. Later, the implications of our results for prominence seismology are discussed in Section~\ref{sec:seis}. Finally, the  conclusions of this work are given in Section~\ref{sec:conc}.

\section{Model}
\label{sec:model}

\begin{figure*}[!htp]
\centering
\includegraphics[width=1.75\columnwidth]{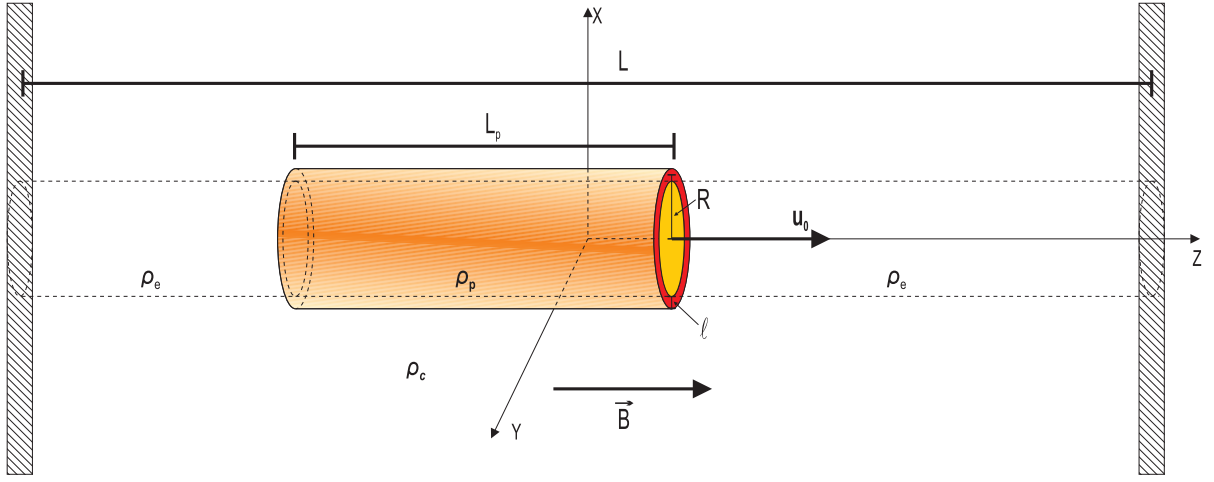}
\caption{Sketch of the prominence thread model used in this work. \label{fig:model}}
\end{figure*}

The prominence thread model used in this paper is schematically shown in Figure~\ref{fig:model}. It is composed of a straight and cylindrical magnetic flux tube of radius $R$ and length $L$.  The $z$-axis is chosen so that it coincides with the axis of the tube.  The magnetic field is ${\bf B} = B \hat{e}_z$, with $B$ constant everywhere. The tube ends are located at $z= \pm L/2$ and are fixed by two rigid walls representing the solar photosphere. The magnetic tube is partially filled with prominence plasma, i.e., the prominence thread, with length $\lp$ and density $\rp$, while the rest of the tube, i.e., the evacuated part, is occupied by less dense plasma of density $\re$. The external plasma represents the coronal medium with density $\rc$. According to the observed typical values of thread widths and lengths from the high-resolution observations of prominences \citep[see, e.g.,][]{lin2011}, the ranges of realistic values of $R$ and $\lp$ are 50~km~$\lesssim R \lesssim$~300~km and 3,000~km~$\lesssim \lp \lesssim$~28,000~km. The total tube length, $L$, is very difficult to measure from the observations, but one can relate $L$ to the typical spatial scale in prominences and filaments, i.e., $L \sim 10^5$~km. These estimations of $L$, $\lp$, and $R$ imply that prominence threads are very thin and long structures. The ratio of the prominence density to the coronal density is a large parameter. The value $\rp / \rc = 200$ is usually considered in the literature, although we do not fix the density contrast at the present stage. For simplicity we assume that the evacuated part of the tube has the same density as the corona so that we take $\re = \rc$. 

In addition, in the prominence thread we consider a transversely inhomogeneous transitional layer of thickness $l$, that continuously connects the internal prominence density, $\rp$, to the external coronal density, $\rc$. The limits $l/R = 0$ and $l/R=2$ correspond to a thread without transverse transitional layer and a fully inhomogeneous thread in the radial direction, respectively. The present paper adopts the $\beta=0$ approximation, where $\beta$ refers to the ratio of the gas pressure to the magnetic pressure. Hence, the plasma temperature is irrelevant for this study and we can choose the density profile in the transitional layer arbitrarily. The presence of this transitional layer introduces damping of the kink waves by resonant absorption \citep[see, e.g.,][]{goossens92,goossens2006,goossens2011rev}. Finally, as a key ingredient of the model, there is a flow along the tube with a constant velocity $u_0$. As a consequence, the prominence thread moves along the tube as a block at the velocity $u_0$.

Standing kink waves supported by the present model have been investigated in the past under several simplifications. In the absence of flow and for $l/R = 0$, there are the studies by \citet{joarder97} and \citet{diaz2001} in Cartesian geometry, and by \citet{diaz2002} in cylindrical geometry. These works considered arbitrary values of $R$ and $L$, which involves a complicated mathematical formalism. \citet{dymovaruderman} derived a more simplified approach valid for thin cylindrical tubes, i.e., $R/L \ll 1$. The thin tube approximation provides very accurate results when compared to the general solution. Subsequently, the case $l/R \neq 0$ in the absence of flow was addressed by \citet{dymovaruderman2} and \citet{solerstatic} again in the thin tube approximation, and by \citet{arregui2d} in the general case. In the presence of flow, the problem was studied  by  \citet{terradas2008} numerically and  by \citet{solerthreadflow} combining analytic and numerical methods. However, both works considered $l/R =0$ and so  resonant damping was absent from their investigations.

\section{Mathematical method}
\label{sec:maths}

The mathematical method followed in this work is based on the theory developed by \citet{ruderman2011a,ruderman2011b}, who derived the governing equation describing standing kink waves in the thin tube and thin boundary approximations \citep[see the Equation~(23) of][]{ruderman2011b}. We refer the reader to the original paper by \citet{ruderman2011b} for a detailed derivation of the governing equation. The thin tube approximation means that we are restricted to the case $R/L \ll 1$ and $R/\lp \ll 1$. This is not a problem since the realistic values of $L$, $\lp$, and $R$ satisfy these constraints. The thin boundary approximation means that we take $l/R \ll 1$, i.e., an abrupt but continuous transition in density in the transverse direction.  

Equation~(23) of \citet{ruderman2011b} can be solved analytically using the Wentzel-Kramers-Brillouin (WKB) approximation \citep[see, e.g.,][]{bender}. The WKB approximation is appropriate when the background properties are slowly varying functions of space and/or time. In the present application of the WKB approximation we assume that the time scale related to the waves, e.g., the period, is much shorter than the time scale related to the changes of the background configuration. As in \citet{solerthreadflow} we define the parameter $\delta$ as
\begin{equation}
 \delta \equiv \frac{u_0}{L}, \label{eq:deltadef}
\end{equation}
with $u_0$ the flow velocity and $L$ the total length of the tube. The validity of the WKB approximation is restricted to small values of  $\delta$ so that $P \delta  \ll 1$, where $P$ is the period of the standing kink wave. In the observations by \citet{okamoto}, the mean flow velocity and period are $u_0 \approx 40$~km~s$^{-1}$ and $P\approx 3$~min. For $L \sim 10^5$~km these values give $P \delta  \approx 0.072$. Hence the condition of applicability of the WKB approximation is fulfilled in the present case. Using the WKB approximation, \citet{ruderman2011b} arrives at the expression for the transverse displacement of the fundamental mode of the magnetic tube, $\eta (t,z)$, namely
\begin{equation}
\eta (t,z) = A(t) W_0(t,z) \exp \left[ i F(t,z) \right] \exp \left[ i \int_0^t \omega(\tau) {\rm d}\tau\right],
\end{equation}
where $ A(t) $, $W_0(t,z)$, and $F(t,z) $ are real functions, and $\omega(t)$ is the instantaneous frequency at time $t$. The function $A(t)$ is positive and represents the oscillation amplitude as a function of time. The function $F(t,z)$ represents a small phase shift that can be neglected \citep[see][]{ruderman2011b} and, therefore, it is omitted hereafter. The function $W_0(t,z)$ satisfies the equation
\begin{equation}
\vk^2 \frac{\partial^2 W_0}{\partial z^2} + \omega^2 W_0 = 0, \label{eq:w0}
\end{equation}
along with the boundary conditions $W_0 (t,  \pm L/ 2 ) = 0$ due to photospheric line-tying. The function $W_0(t,z)$ is defined so that $W_0(t,z) > 0$ and $\max\left[W_0(t,z) \right] =  1$. In Equation~(\ref{eq:w0}) $\vk$ is the kink velocity, which in our model is defined as
\begin{equation}
  \vk(z,t) = \left\{ \begin{array}{lll}
                     \vkp, & \textrm{if} & z_- \leq z \leq z_+, \\
		      \vke, & &\textrm{otherwise}, 
                    \end{array} \right. \label{eq:kinkspeed}
\end{equation}
where
\begin{equation}
  \vkp = \sqrt{\frac{2 B^2}{\mu_0 \left( \rp + \rc \right)}}, \qquad \vke = \sqrt{\frac{B^2}{\mu_0 \rc }},
\end{equation}
with $\mu_0$ the magnetic permeability of free space and
\begin{equation}
 z_\pm = z_0 +u_0 t \pm \frac{\lp}{2}, 
\end{equation}
with $z_0$ corresponding to the position of the center of the prominence thread with respect to the center of the magnetic tube at $t=0$. Note that $z_0 < 0$ if the thread is initially located on the left-hand side to the center of the tube, whereas $z_0 > 0$ if the thread is initially located on the right-hand side. We also note that the flow velocity, $u_0$, only explicitly appears in the expressions of $z_+$ and $z_-$. 

The equation governing the amplitude $ A(t) $ corresponds to Equation~(66) of \citet{ruderman2011b}, namely
\begin{equation}
 \frac{\der}{\der t} \left( \omega I A^2  \right) = -\Gamma A^2, \label{eq:amp}
\end{equation}
where
\begin{eqnarray}
 I &=& \int_{-L/2}^{L/2} \left( \rho_{\rm i} + \rc \right) W_0^2 \der z, \label{eq:i} \\
\Gamma &=& \frac{\pi}{2} \frac{\mu}{B^2} \frac{\omega^4}{R} \sum_{n=1}^N \int_{-L/2}^{L/2} \left( \rho_{\rm i} - \rc \right) W_0 w_n(r_n) \der z \nonumber \\
 & & \times \int_{-L/2}^{L/2} \frac{ \rho_{\rm t}(r_n) - \rc}{\left| \Delta_n \right|} W_0 w_n(r_n) \der z. \label{eq:gamma}
\end{eqnarray}
In the expressions of $I$ and $\Gamma$, $\rho_{\rm i}$ is the internal density given by
\begin{equation}
  \rho_{\rm i} = \left\{ \begin{array}{lll}
                     \rp, & \textrm{if} & z_- \leq z \leq z_+, \\
		      \rc, & &\textrm{otherwise}, 
                    \end{array} \right. 
\end{equation}
and $\rho_{\rm t}(r)$ is the density in the transitional layer, with $\rho_{\rm t}(r_n)$ the corresponding value at the n-th resonant position $r=r_n$. The function $w_n$ is the eigenfunction of the Alfv\'en boundary value problem
\begin{equation}
\va^2(r) \frac{\partial^2 w_n(r)}{\partial z^2} + \omega^2_{\rm A,n}(r) w_n(r) = 0, \label{eq:wn}
\end{equation}
along with $w_n(r) = 0$ at $z = \pm L/2$, where $\va^2(r) = B^2 / \mu  \rho_{\rm t}(r)$ is the local Alfv\'en velocity squared. In Equation~(\ref{eq:wn})  $\omega_{\rm A,n}(r)$ is the n-th Alfv\'en eigenvalue. The eigenfunction $w_n$ satisfies the normalization condition
\begin{equation}
 \int_{-L/2}^{L/2} \frac{1}{\va^2}w_nw_m \der z = \delta_{mn},\label{eq:normwn}
\end{equation}
where  $\delta_{mn}$ is the Kronecker delta. The position of the n-th resonance, $r_n$, is given by the condition $\omega^2 = \omega^2_{\rm A,n}(r_n)$. Finally, the quantity $\Delta_n$ is given by
\begin{equation}
  \Delta_n = - \left. \frac{\partial \omega^2_{\rm A,n}}{\partial r} \right|_{r_n}.
\end{equation}

We consider the case $l/R = 0$ in order to compare with the expressions used by \citet{solerthreadflow}. In the absence of resonant damping $\Gamma = 0$ and the equation for the amplitude $A(t)$ (Equation~(\ref{eq:amp})) can be expanded to
\begin{equation}
 \frac{\der A(t)}{\der t} + \frac{1}{2\omega} \frac{\der \omega}{\der t} A(t) + \frac{1}{2I} \frac{\der I}{\der t} A(t) = 0.\label{eq:ampexpanded}
\end{equation}
We compare Equation~(\ref{eq:ampexpanded}) to Equation~(20) of \citet{solerthreadflow}. The only difference between the equation for the amplitude used by \citet{solerthreadflow} and the equation derived using the more general method by \citet{ruderman2011b} resides in the term with the temporal derivative of $I$ in Equation~(\ref{eq:ampexpanded}). The equation of \citet{solerthreadflow} does not take into account the dependence of $I$ on time. For the fundamental mode and slow flows, $I$ may  be roughly independent on time. This can be checked once $W_0(t,z)$ is found. In that case, the equation of \citet{solerthreadflow} is an approximate description of the fundamental mode amplitude.

Using the method developed by \citet{ruderman2011a,ruderman2011b} and summarized in the previous paragraphs, the problem of studying damped standing oscillations of flowing threads is reduced to finding the functions $W_0(t,z)$ and $A(t)$, along with the expression for the instantaneous frequency $\omega(t)$. This is done in the following subsections.

\subsection{Finding $W_0(t,z)$ and $\omega(t)$}

To find the function $W_0(t,z)$ we need to solve the boundary value problem defined in Equation~(\ref{eq:w0}) with $W_0 (t,  \pm L/ 2 ) = 0$.  In addition, since the equilibrium is piecewise constant in $z$ we need to provide additional boundary conditions at $z=z_\pm$. Since $z=z_\pm$ correspond to contact discontinuities, the boundary conditions are that $W_0(t,z)$ is continuous at $z = z_\pm$ \citep{goedbloed}. Thus, the process here is equivalent to the process followed by \citet{solerthreadflow} to find their function $Q_1(t,z)$. The reader is referred to \citet{solerthreadflow} for details. The expression for $W_0(t,z)$ is
\begin{equation}
 W_0(t,z) = \left\{ \begin{array}{lll}
             D_1 \sin \left[ \frac{\omega}{\vke} \left( z + \frac{L}{2}  \right) \right] &\textrm{if}& z < z_-, \\
		\cos \left( \frac{\omega}{\vkp} z + \phi \right) &\textrm{if}&  z_- \leq z \leq z_+, \\
	    D_2 \sin \left[ \frac{\omega}{\vke} \left( z - \frac{L}{2}  \right) \right] &\textrm{if}& z > z_+,
            \end{array}
	  \right. \label{eq:expw0}
\end{equation}
where $\phi = \phi(t)$ is a time-dependent phase, and $D_1$ and $D_2$ are given by
\begin{equation}
 D_1 = \frac{\cos \left( \frac{\omega}{\vkp} z_- + \phi \right)}{\sin \left[ \frac{\omega}{\vke} \left( z_- + \frac{L}{2}  \right) \right]}, \qquad D_2 = \frac{\cos \left( \frac{\omega}{\vkp} z_+ + \phi \right)}{\sin \left[ \frac{\omega}{\vke} \left( z_+ - \frac{L}{2}  \right) \right]}.
\end{equation}
The expression for $W_0(t,z)$ takes into account the condition $\max\left[W_0(t,z) \right] =  1$.

The time-dependent frequency, $\omega(t)$, requires to obtain first the time-dependent dispersion relation by providing additional boundary conditions which enable us to eliminate the time-dependent phase, $\phi$. These additional conditions are that $\partial W_0 / \partial z$ is continuous at $z=z_\pm$. The expression for $\phi$ is
\begin{equation}
 \phi = - \frac{\omega}{\vkp} z_- - \arctan \left\{ \frac{\vkp}{\vke} \cot \left[\frac{\omega}{\vke}\left( z_- + \frac{L}{2} \right)  \right] \right\},
\end{equation}
and the dispersion relation is Equation~(14) of \citet{solerthreadflow}. The fundamental mode corresponds to the solution with the lowest frequency. By performing the first-order expansion of their Equation~(14) with respect to the small parameter $\rc/\rp$, \citet{solerthreadflow} obtained an approximate expression for $\omega(t)$ (their Equation~(15)), which in the limit $\rc /\rp \ll 1$ realistic of prominences reduces to
\begin{equation}
 \omega \left( t\right) \approx   \frac{2 \vkp \sqrt{L/\lp}}{\sqrt{ \left( L - \lp \right) \left( L + \frac{1}{3} \lp \right) - 4 \left( z_0 + u_0 t \right)^2  } }. \label{eq:freq} 
\end{equation}
This is the expression for $\omega \left( t\right)$ used in the following computations. In the absence of flow, $u_0 = 0$ and $\omega$ becomes independent of time and corresponds to the normal mode frequency \citep[see][]{solerstatic}.

Since $W_0(t,z)$ is known, we can compute the integral $I$ (Equation~(\ref{eq:i})). The result is
\begin{eqnarray}
 I &=& 2 \rc D_1^2 \left[ \frac{z_- + L/2}{2} - \frac{\vke}{4 \omega} \sin \left( \frac{2 \omega}{\vke} (z_- + L/2) \right) \right] \nonumber \\
&-& 2 \rc D_2^2 \left[ \frac{z_+ - L/2}{2} - \frac{\vke}{4 \omega} \sin \left( \frac{2 \omega}{\vke} (z_+ - L/2) \right) \right] \nonumber \\
&+& (\rp+\rc) \left\{ \frac{z_+ - z_-}{2} + \frac{\vkp}{4 \omega} \left[ \sin\left( \frac{2 \omega}{\vkp}z_+ + \phi \right) \right. \right. \nonumber \\
 &-& \left. \left. \sin\left( \frac{2 \omega}{\vkp}z_- + \phi \right) \right] \right\}. \label{eq:icomp}
\end{eqnarray}
As before, we take advantage of the fact that the fundamental mode is the solution with the lowest frequency. We perform the  first-order expansion of Equation~(\ref{eq:icomp}) with respect to the small parameter $\rc/\rp$ to get the approximate expression of $I$ valid for the fundamental mode, namely
\begin{equation}
 I \approx \left( \rp + \rc \right)\left(z_+ - z_- \right) = \left( \rp + \rc \right) \lp.\label{eq:icomp2}
\end{equation}
Thus, we find that the integral $I$ is time-independent in this approximation. We can now check the equation for the amplitude used by \citet{solerthreadflow} in the absence of resonant damping, i.e., $l/R=0$. When $I$ is constant in time,  Equation~(\ref{eq:ampexpanded}) reverts to Equation~(20) of \citet{solerthreadflow}. Therefore, the equation for the amplitude used by \citet{solerthreadflow} is justified in this approximation.

With the use of the approximate expression for $I$ given in Equation~(\ref{eq:icomp2}), we obtain the general equation for the amplitude when $l/R \neq 0$ (Equation~(\ref{eq:amp})), namely
\begin{equation}
 \frac{\der}{\der t} \left( \omega A^2  \right) = -\frac{\Gamma}{\left( \rp + \rc \right) \lp} A^2. \label{eq:amp2}
\end{equation}
The next step is to compute $\Gamma$ in order to integrate Equation~(\ref{eq:amp2}).

\subsection{Finding $\Gamma$}

Equation~(\ref{eq:gamma}) for $\Gamma$ takes into account the possibility of multiple resonances. From hereon we assume that there is only one resonance position at $r=r_1$, and we set $N=1$ in Equation~(\ref{eq:gamma}). This is the same assumption as in \citet[Section~5.1]{ruderman2011b}. This assumption was also used by \citet{solerstatic} and was later checked to be correct by \citet{arregui2d}. In addition, we take into account that the density in the evacuated part of the tube, $\re$, is set equal to the coronal density, $\rc$, so that $l/R = 0$ in the evacuated region. Thus, we can change the limits of integration in Equation~(\ref{eq:gamma}) from $[-L/2, L/2]$ to $[z_-, z_+]$. Also, since $\rp$, $\rc$, and $\rho_{\rm t}(r_1)$ are constants, we can move them out of the integrals. The same is true for $\left| \Delta_1 \right|$ since this quantity is independent of $z$. Finally, the expression for $\Gamma$ becomes
\begin{equation}
 \Gamma = \frac{\pi}{2} \frac{\mu}{B^2} \frac{\omega^4}{R} \frac{( \rp - \rc )( \rho_{\rm t}(r_1) - \rc )}{\left| \Delta_1 \right|} \Lambda^2, \label{eq:gamma2}
\end{equation}
with
\begin{equation}
 \Lambda = \int_{z_-}^{z_+} W_0 w_1(r_1) \der z \label{eq:lamdba}
\end{equation}

At the resonance position $r=r_1$, so  $\omega^2 = \omega^2_{\rm A,1}(r_1)$. By comparing Equations~(\ref{eq:w0}) and (\ref{eq:wn}) we see that $W_0$ and $w_n(r_1)$ are the eigenfunctions of the same eigenvalue problem when
\begin{equation}
 \rho_{\rm t}(r_1) = \frac{\rp + \rc}{2}. \label{eq:rt}
\end{equation}
So, we can put $w_n(r_1)$ proportional to $W_0$, namely $w_n(r_1) = Q W_0$, where $Q$ is a constant of proportionality, and Equation~(\ref{eq:lamdba}) becomes
\begin{eqnarray}
 \Lambda = Q \int_{z_-}^{z_+} W_0^2 \der z &=& Q \left\{ \frac{z_+ - z_-}{2} + \frac{\vkp}{2 \omega} \left[ \sin \left( \frac{2\omega}{\vkp}z_+ + 2\phi \right) \right. \right. \nonumber \\
 &-& \left.  \left. \sin \left( \frac{2\omega}{\vkp}z_- + 2\phi \right) \right] \right\}. \label{eq:lamdba2}
\end{eqnarray}
As before, we only take into account the fundamental mode and perform the first-order expansion of Equation~(\ref{eq:lamdba2}) with respect to $\rc/\rp$. We also use the normalization condition (Equation~(\ref{eq:normwn})) with $m=n=1$ to find that $Q \approx \vkp / \lp^{1/2}$ in this  approximation for the fundamental mode. Thus, we obtain the approximate $\Lambda$ as
\begin{equation}
 \Lambda \approx \vkp \lp^{1/2}.\label{eq:lamdba3}
\end{equation}

Now we compute $\left| \Delta_1 \right|$. To do so we use again the resonant condition $\omega^2 = \omega^2_{\rm A,1}(r_1)$ and Equation~(\ref{eq:rt}) to write
\begin{equation}
 \omega^2_{\rm A,1}(r) = \omega^2 \frac{\va^2(r)}{\vkp^2} = \omega^2\frac{\rp+\rc}{2\rho_{\rm t}(r)}.
\end{equation}
Hence the expression for $\left| \Delta_1 \right|$ is
\begin{equation}
 \left| \Delta_1 \right| = \omega^2 \frac{2}{\rp + \rc} \left| \frac{\partial \rho_{\rm t}}{\partial r} \right|_{r_1}. 
\end{equation}
We next write 
\begin{equation}
 \left|\frac{\partial \rho_{\rm t}}{\partial r}\right|_{r_1} = F \frac{\pi^2}{4} \frac{\rp - \rc}{l},
\end{equation}
where $F$ is a factor that depends on the form of the transverse density profile. For example, $F=4/\pi^2$ for a linear profile and $F=2/\pi$ for a sinusoidal profile with $r_1 = R$. The expression for $\left| \Delta_1 \right|$ becomes
\begin{equation}
  \left| \Delta_1 \right| = F \frac{\pi^2}{2} \frac{\omega^2}{l}   \frac{\rp - \rc}{\rp + \rc}. 
\end{equation}
Finally, we substitute all these results in Equation~(\ref{eq:gamma2}) and arrive at the expression for $\Gamma$, namely
\begin{equation}
 \Gamma = \frac{ \omega^2}{\pi F}   \frac{l}{R} \left(\rp - \rc  \right) \lp. \label{eq:gamma3}
\end{equation}

\subsection{Time-dependent amplitude $A(t)$}

We substitute Equation~(\ref{eq:gamma3}) in Equation~(\ref{eq:amp2}) to write the equation for the time-dependent amplitude $A(t)$ as
\begin{equation}
 \frac{\der A}{\der t} = - \left( \gamma \omega + \frac{1}{2\omega} \frac{\der \omega}{\der t} \right)A, \label{eq:amp3}
\end{equation}
with
\begin{equation}
 \gamma = \frac{1}{2\pi F}   \frac{l}{R} \frac{\rp-\rc}{\rp+\rc}.
\end{equation}
Equation~(\ref{eq:amp3}) can be easily integrated using the expression for $\omega(t)$ given in Equation~(\ref{eq:freq}). The result is
\begin{equation}
 A(t) = A_0  \left[ \frac{\left( L - \lp \right) \left( L + \frac{1}{3} \lp \right) - 4 \left( z_0 + u_0 t \right)^2}{\left( L - \lp \right) \left( L + \frac{1}{3} \lp \right) - 4 z_0^2} \right]^{1/4}  e^{ - \gamma \Omega(t)},\label{eq:amp4}
\end{equation}
with the function $\Omega(t)$ given by
\begin{eqnarray}
 \Omega(t) &=& \frac{\vkp}{u_0} \sqrt{\frac{L}{\lp}} \left[ \arctan\left( \frac{2(z_0+u_0 t)}{\sqrt{ \left( L - \lp \right) \left( L + \frac{1}{3} \lp \right) - 4 \left( z_0 + u_0 t \right)^2  }} \right) \right. \nonumber \\ 
&-& \left.  \arctan\left( \frac{2z_0}{\sqrt{ \left( L - \lp \right) \left( L + \frac{1}{3} \lp \right) - 4 z_0^2  }} \right)  \right].
\end{eqnarray}

The first factor on the right-hand side of Equation~(\ref{eq:amp4}), i.e., $A_0$, represents the initial amplitude at $t=0$. The second factor on the right-hand side of Equation~(\ref{eq:amp4}), i.e., that with the square brackets, is the same as that found in Equation~(22) of \citet{solerthreadflow}. This term accounts for the change of the amplitude due to the motion of the prominence thread along the magnetic tube. This contribution can cause either amplification or damping depending on the instantaneous position of the thread with respect to the center of the magnetic tube. This term is equal to unity in the static case, i.e., for $u_0 = 0$. The third factor on the right-hand side of Equation~(\ref{eq:amp4}) is an exponential factor that accounts for  damping due to resonant absorption. 

To recover the result in the static case without flow, we evaluate the limit of  $\Omega(t)$ when $u_0 \to 0$, namely
\begin{equation}
 \lim_{u_0 \to 0}\Omega(t) = \omega_0 t,
\end{equation}
where $\omega_0$ the frequency given in Equation~(\ref{eq:freq}) with $u_0 = 0$. In that case, the exponential factor on the right-hand side of Equation~(\ref{eq:amp4}) becomes $\exp \left( -\gamma \omega_0 t \right)$, so that we consistently revert to the static case studied by \citet{solerstatic} and \citet{arregui2d}.

\section{Parametric study}

\label{sec:param}

Here we explore the impact of the various model parameters on the amplitude. In particular we focus on the effects of the thickness of the transitional layer, $l/R$, and the flow velocity, $u_0$. We assume a linear variation for the density in the transitional layer.

 \begin{figure}[!htp]
\centering
\includegraphics[width=.85\columnwidth]{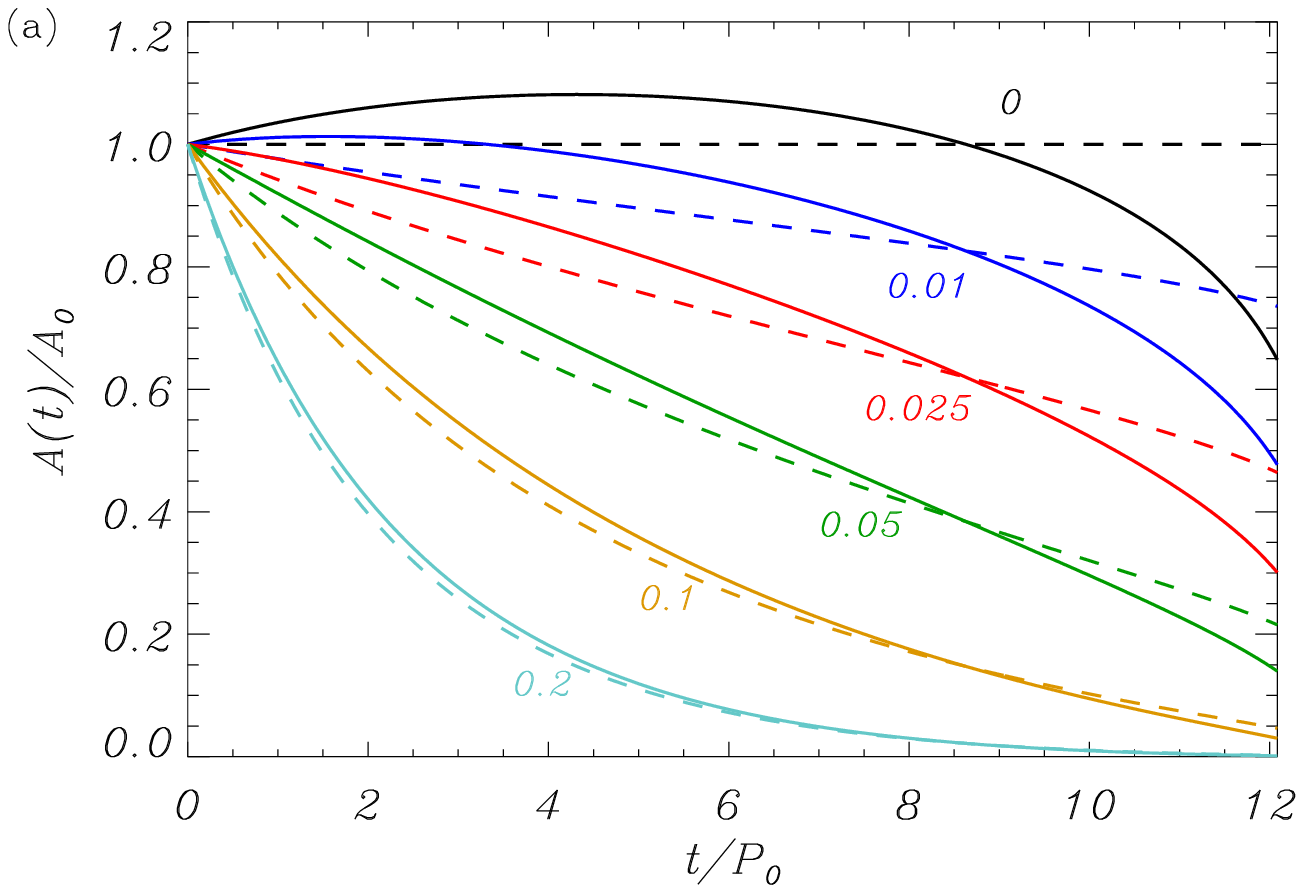}
\includegraphics[width=.85\columnwidth]{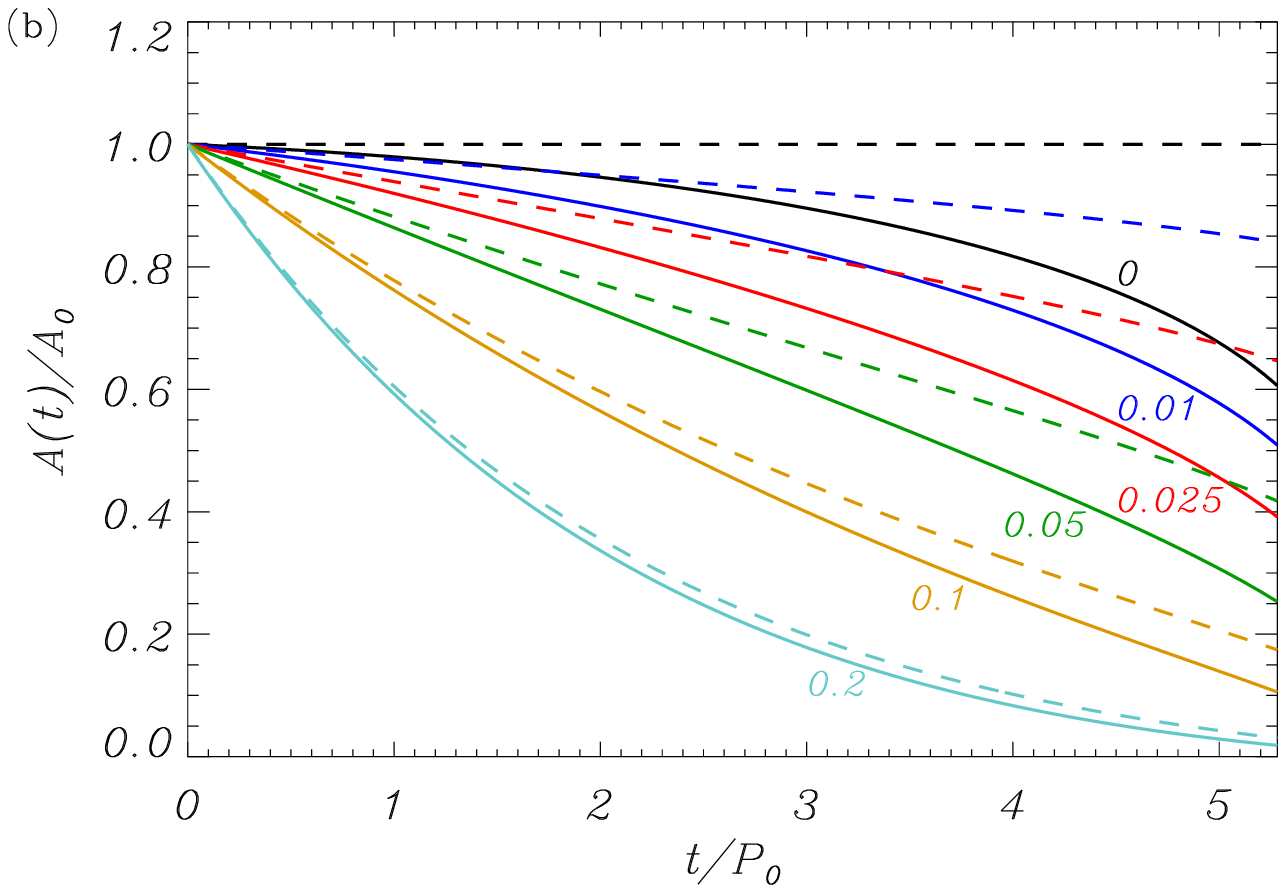}
\caption{$A(t)/A_0$ as a function of $t/P_0$, with $P_0$ the instantaneous period at $t=0$. The numbers next to the various lines indicate the value of $l/R$ used. The solid line is the complete result, while the dashed line is the equivalent result but with $u_0 = 0$. We have used ($a$) $z_0/L = -0.25$ and ($b$) $z_0 / L = 0.1$.  In all computations $\lp/L = 0.1$, $u_0 / \vap = 0.1$, and $\rp / \rc = 200$.   \label{fig:amp}}
\end{figure}

Figure~\ref{fig:amp} shows  $A(t)/A_0$ as a function of $t/P_0$, with $P_0=2\pi/\omega_0$ the instantaneous period at $t=0$. We plot the results for several values of $l/R$ and $z_0/L$, and compare the amplitude obtained in the presence and in the absence of flow. We find that the amplitude in the presence of flow is substantially different from that for $u_0 = 0$ when the transitional layer is very thin, so that damping due to resonant absorption is very weak. In this case the variation of the amplitude is mainly governed by the effect of the flow, described by the factor with the square brackets on the right-hand side of Equation~(\ref{eq:amp4}).  It is possible to obtain amplification of the amplitude at short times when $z_0/L <0$ and the transitional layer is very thin (top lines in Fig.~\ref{fig:amp}a). As $l/R$ increases, the damping term, which is the exponential factor on the right-hand side of Equation~(\ref{eq:amp4}), becomes more important and the amplitude is almost the same in both cases $u_0 = 0$ and $u_0 \neq 0$, i.e., the effect of the flow is then of minor importance. When $z_0/L >0$ (Fig.~\ref{fig:amp}b) amplification is not possible regardless the value of $l/R$ and damping is always observed.

\begin{figure}[!htp]
\centering
\includegraphics[width=.85\columnwidth]{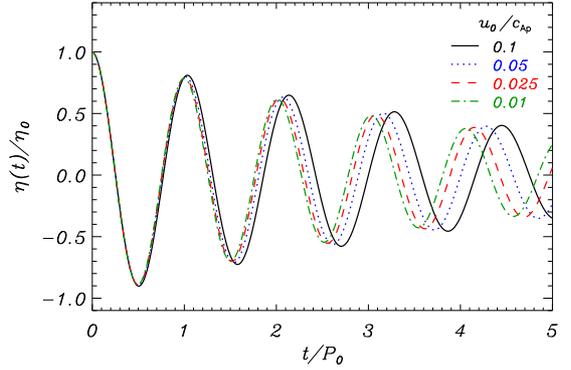}
\caption{Prominence thread transverse displacement, $\eta(t)/\eta_0$, as a function of $t/P_0$, with $P_0$ the instantaneous period at $t=0$, for various values of $u_0 / \vap$ indicated within the figure. In all computations $\lp/L = 0.1$, $l/R = 0.1$, $z_0/L = -0.25$, and $\rp / \rc = 200$.   \label{fig:amp2}}
\end{figure}

Figure~\ref{fig:amp2} displays the thread displacement as a function of $t/P_0$ for different values of the flow velocity (the remaining parameters are indicated in the caption of the Figure). We obtain a progressive phase shift between the solutions corresponding to different velocities. This is so because the instantaneous period, $P(t)=2\pi/\omega(t)$, is a function of the flow velocity according to Equation~(\ref{eq:freq}). However, for the set of parameters used in Figure~\ref{fig:amp2} the amplitude is weakly affected by the flow velocity. In particular, the value $l/R = 0.1$ used in Figure~\ref{fig:amp2}  is large enough for the amplitude to be dominated by resonant damping.

\section{Implications for prominence seismology}

\label{sec:seis}

The results presented in this paper have implications for seismology of solar prominences \citep[see][]{arregui12a}. Adopting resonant absorption as damping mechanism, the inversion technique of  \citet{goossens08} has been applied to the case of prominence thread oscillations \citep[see][]{arreguiballester,solerstatic}. Due to the high density contrast of the prominence plasma with respect to the coronal plasma, a direct estimation of $l/R$ is possible when measures of both period and damping time are available. The value of $l/R$ inferred by seismology may be used to test thread models, because the transverse inhomogeneity length scale is a crucial parameter for the energy balance of the prominence threads with the surrounding hot coronal plasma \citep[see, e.g.,][]{pojoga,cirigliano,labrossereview}. The analytic equation used to infer $l/R$ is \citep{goossens08} 
\begin{equation}
\frac{l}{R} = F \frac{P}{\tau_{\rm D}} \frac{\rp + \rc}{\rp - \rc} \approx F \frac{P}{\tau_{\rm D}}, \label{eq:lrseis}
\end{equation}
where $P$ and $\tau_{\rm D}$ are the (constant) period and the damping time obtained after fitting  to the observed oscillation a harmonic function with amplitude proportional to $\exp \left( - t/\tau_{\rm D} \right)$. This expression does not take the flow of the prominence thread into account. In this Section we test the robustness of seismological estimates when flow is present and Equation~(\ref{eq:lrseis}) is used in combination with typical fitting methods for the oscillation amplitude. In addition, the influence of background noise is also studied.  

\subsection{Example}

\begin{figure}[!htp]
\centering
\includegraphics[width=.75\columnwidth]{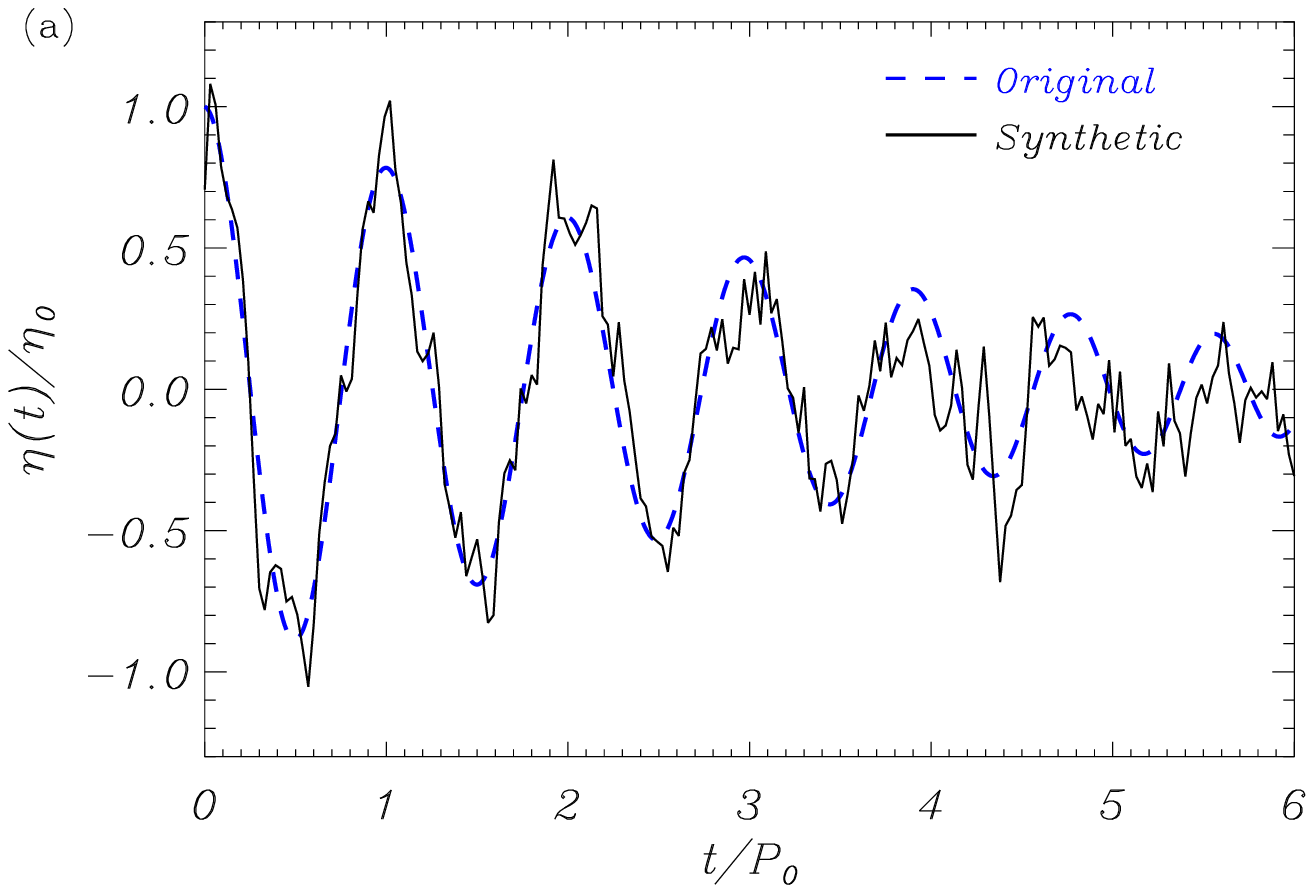}
\includegraphics[width=.75\columnwidth]{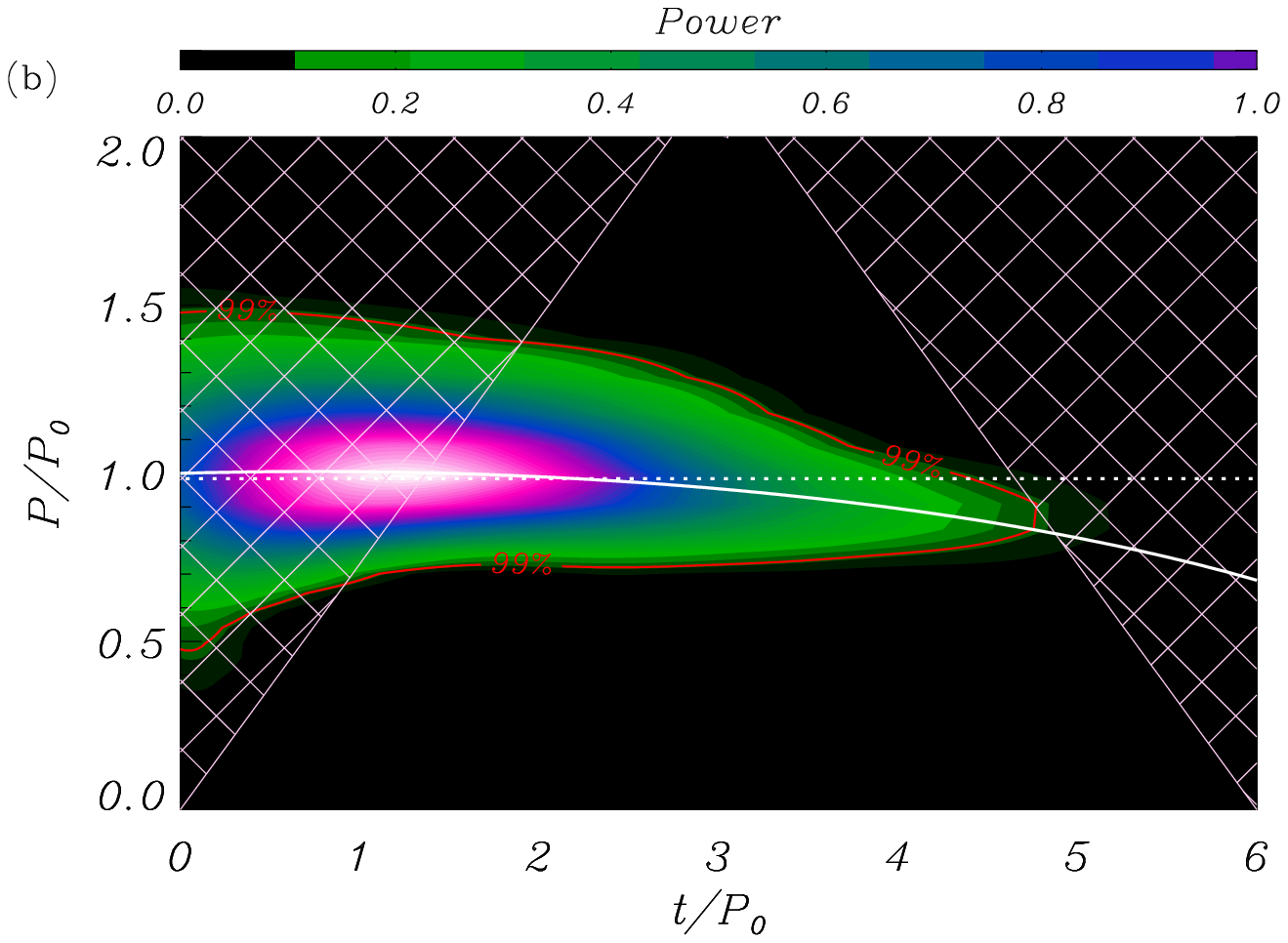}
\includegraphics[width=.75\columnwidth]{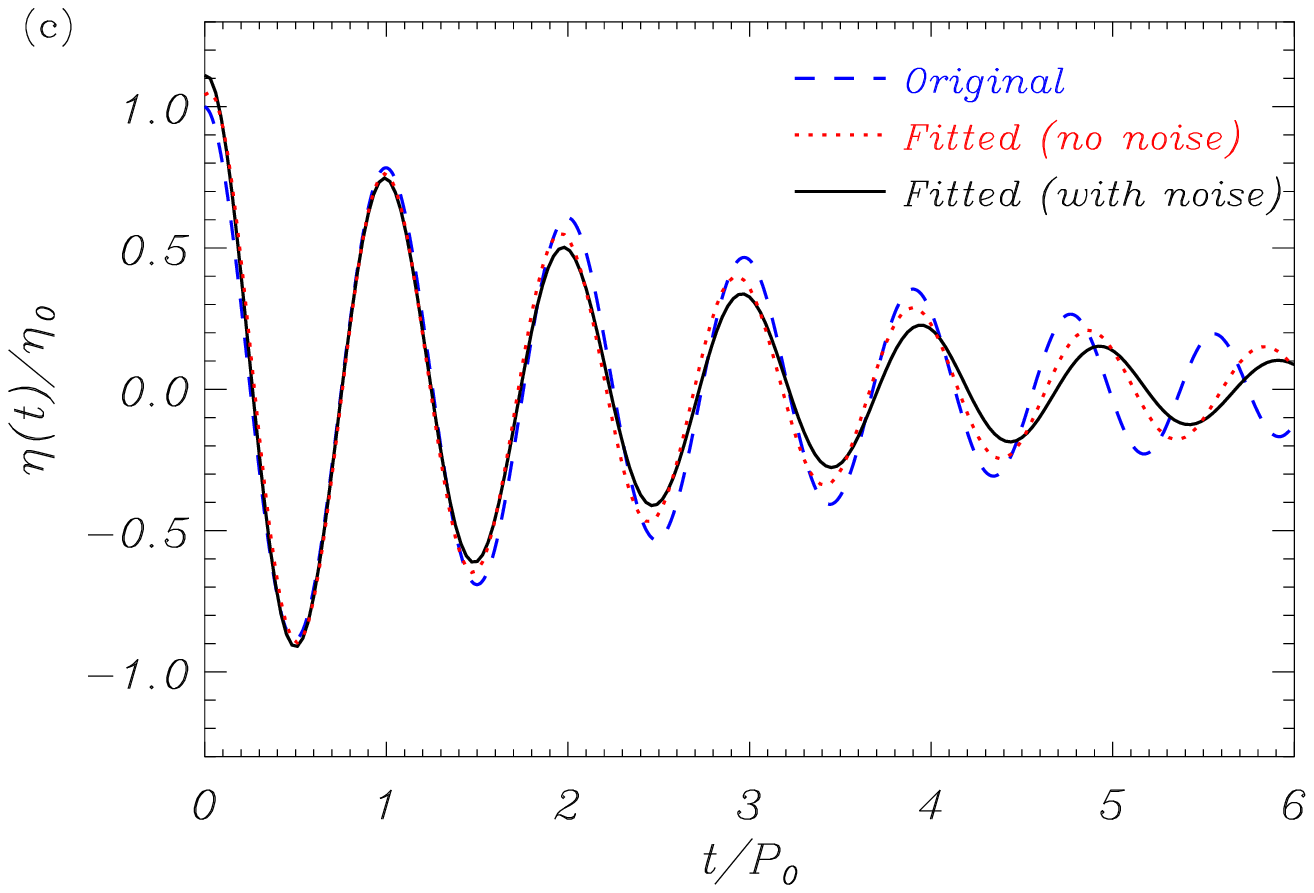}
\caption{(a) Original (dashed) and synthetic (solid) data used in the seismological test. (b) Wavelet power spectrum for the dimensionless period, $P / P_0$, corresponding to the synthetic signal displayed in panel (a). The white solid line is the original data instantaneous period, whereas the horizontal dotted line is the period obtained from the fitting method. The red solid line denotes 99\% of confidence level. (c) Comparison of the original (dashed) and fitted signals with noise (solid) and without noise (dotted). \label{fig:seis}}
\end{figure}

We synthetically generate a signal aiming to represent a prominence thread transverse oscillation detected with a real instrument. We use the following set of parameters: $\lp/L = 0.1$, $l/R = 0.1$, $z_0/L = -0.05$, $\rp / \rc = 200$, and $u_0 / \vap = 0.1$. A linear density profile in the transitional layer is used. The corresponding theoretical transverse displacement is displayed in Figure~\ref{fig:seis}(a) with a dashed line. The time series corresponds roughly to 6 oscillation periods. This is approximately the number of periods observed in time series of real events \citep[see, e.g.,][]{ning}. To represent the limited cadence of the instrument we perform a temporal sampling of this signal using, approximately, 33 points per period. For a typical period of 3~min, this corresponds to a cadence of about 5~s. This cadence is of the same order as that used in recent observations \citep[see, e.g.,][]{lin09}. Later, we add to the sampled signal a  randomly generated noise. In this way we try to account for the contribution of the background to the total signal and/or to consider unspecified instrumental uncertainties. Noise is generated with the IDL function \verb&randomu&. Here, we specify the amplitude of the background noise as percentage  of the signal amplitude at $t=0$, namely
\begin{equation}
 \left(\%\right)_{\rm noise} = \frac{\eta_{\rm noise}}{\eta_0}\times 100,
\end{equation}
where $\eta_{\rm noise}$ is the noise amplitude and $\eta_0$ is the signal amplitude at $t=0$. It is more frequent in observational astrophysics to use the signal-to-noise ratio, $S/N$, which is defined as the power ratio between the signal and the background noise. Assuming that the power is proportional to the square of the amplitude, the relation between $\left(\%\right)_{\rm noise}$ and $S/N$ is
\begin{equation}
 S/N = \left( \frac{100}{\left(\%\right)_{\rm noise}} \right)^2.
\end{equation}
For example, $\left(\%\right)_{\rm noise} = 10\%$ corresponds to $S/N = 100$. In the present example we use $\left(\%\right)_{\rm noise} = 50\%$, which is equivalent to $S/N = 4$. Finally, a smoothing algorithm is applied to the noisy data. The resulting signal is plotted in Figure~\ref{fig:seis}a with the solid line. Since the added noise is randomly generated, we can obtain different synthetic signals depending on the seed used in the random number generator. The synthetic data displayed in Figure~\ref{fig:seis}(a) is the one used in this particular example.

 Figure~\ref{fig:seis}(b) shows a wavelet power spectrum \citep{wavelet} of the synthetic signal displayed in Figure~\ref{fig:seis}(a). For comparison we overplot the instantaneous period of the original data. We see that the wavelet spectrum recovers well the period of the original data. However, since the effect of the flow on the period is not very strong and the duration of the time series is limited to 6 periods, the variation of the period with time is not evident in the wavelet spectrum. The added noise also makes difficult the analysis near the end of the time series when the amplitude is low. Based on this result, we believe that the effect of the flow on the period would probably go unnoticed for an observer using wavelet analysis.

We turn again to the synthetic signal of Figure~\ref{fig:seis}(a) and try a different approach. We fit to the synthetic data a typical exponentially damped harmonic function as 
\begin{equation}
f(t) = a_1 \cos \left( a_2 t + a_3 \right) \exp \left( - a_4 t \right), \label{eq:fit}
\end{equation}
where $a_1$, $a_2$, $a_3$, and $a_4$ are four parameters to fit. The fitted period, $P$, and damping time, $\tau_{\rm D}$, are
\begin{equation}
P = \frac{2\pi}{a_2}, \qquad \tau_{\rm D} = \frac{1}{a_4}.
\end{equation}
The fit is made using the IDL function \verb&curvefit&. We perform a fit to the sampled data before and after adding random noise. The result of both fits is shown in Figure~\ref{fig:seis}(c), where we also plot the original curve for comparison. First of all, we obtain a progressive phase shift between the different curves because the fitted function (Equation~(\ref{eq:fit})) does not take into account the variation of the period with time. The constant period obtained from the fit to the noisy data is indicated by the horizontal dotted line in  Figure~\ref{fig:seis}(b). This fitted period agrees well with the maximum in the wavelet spectrum, but overestimates the actual period at later times.

More importantly, we also find different damping rates for the three curves. The damping  of both fitted functions is stronger than the actual damping. In particular, the fitted curve to the noisy data has the strongest damping, i.e., the lowest amplitude. The difference in the damping of the original curve and that of the sampled signal without noise can be attributed to the effect of the flow, whereas in the sampled signal with noise we have to take into account the additional influence of  noise. This has a direct consequence on the seismological estimation of $l/R$ using the fitted $P$ and $\tau_{\rm D}$ in Equation~(\ref{eq:lrseis}). In the original data we used $l/R = 0.1$, while the inferred values are $l/R = 0.13$ for the sampled signal without noise and $l/R = 0.16$ for the sampled signal with noise. We have not computed the uncertainty of the fitted parameters and, therefore, we do not know the relative error on the estimation of $l/R$. However, the results of  this example suggest that the presence of both flow and noise adds significant uncertainty to the determination of $l/R$.

\subsection{Statistical study}

Now, we perform a statistical study of the influence of flow and noise on the determination of $l/R$ and its uncertainty. The purpose here is to know whether the results obtained in the example of the previous subsection are general or, on the contrary, they are particular to this example only. We use the same parameters as in the previous example but vary the value of the flow velocity and the percentage of maximum background noise. 

\begin{figure}[!htp]
\centering
\includegraphics[width=.85\columnwidth]{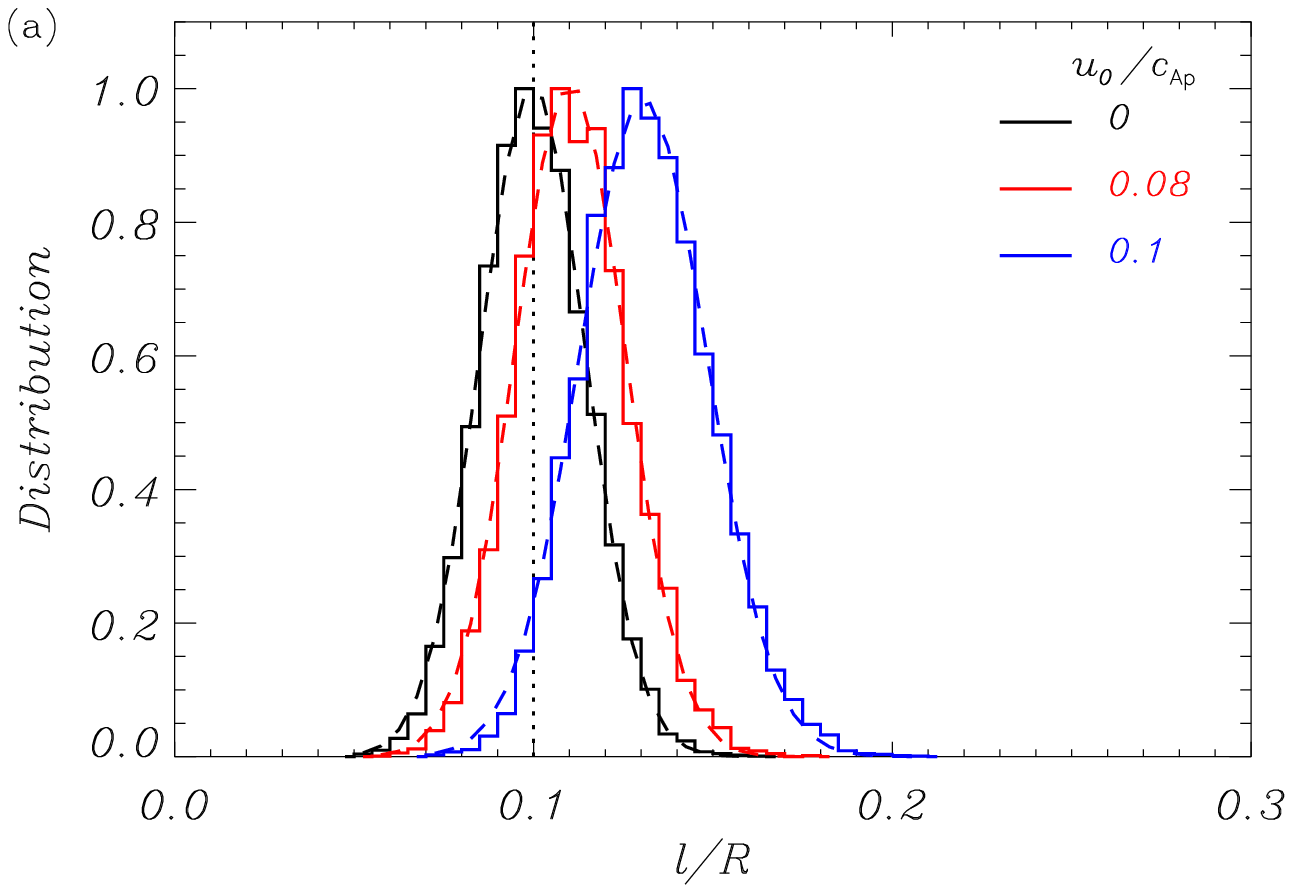}
\includegraphics[width=.85\columnwidth]{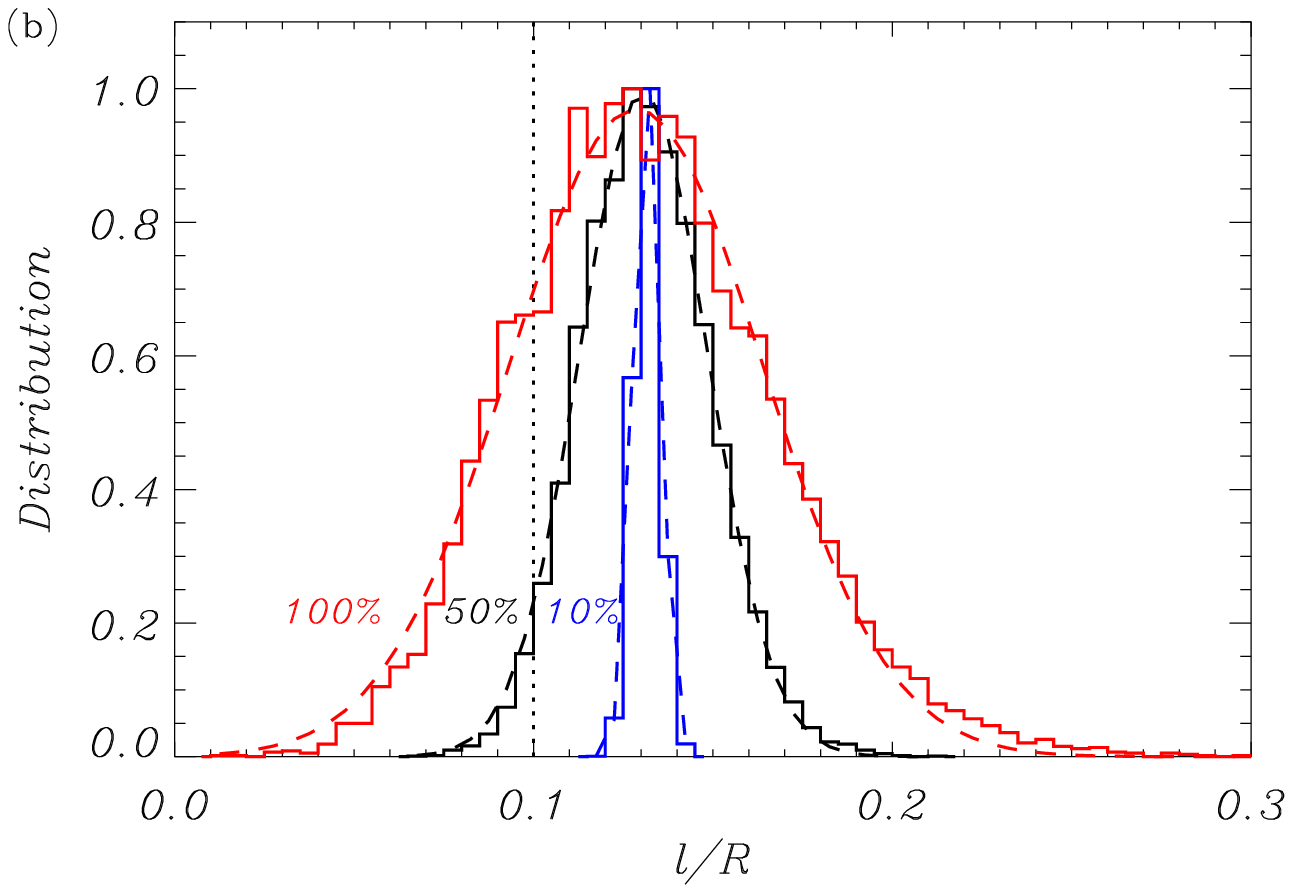}
\caption{(a) Normalized histogram of the seismologically estimated $l/R$ for three  values of the flow velocity (indicated within the panel) and a maximum background noise corresponding to 50\% of the initial amplitude. (b) Same as panel (a) but with $u_0 / \vap = 0.1$ and three  percentages of maximum background noise (indicated next to the lines). In both panels, the dashed lines are the Gaussian fits and the vertical dotted line is the actual $l/R$. \label{fig:stat}}
\end{figure}

First we fix $\left(\%\right)_{\rm noise} = 50\%$ ($S/N = 4$). We follow the procedure explained before to generate a set of $10^4$ synthetic noisy signals for various values of $u_0$. The various synthetic signals are generated using different initial seeds in the random number generator. As in the example, we fit an exponentially damped harmonic function (Equation~(\ref{eq:fit})) to each one of the synthetic signals and infer the value of $l/R$ using Equation~(\ref{eq:lrseis}). Subsequently, we perform a histogram with the full set of $10^4$ estimated $l/R$ for each particular $u_0$. By visual inspection, we determine that the histograms follow the Gaussian distribution. Therefore, we fit to the histograms a Gaussian function of the form
\begin{equation}
g(l/R) = b_0 \exp \left( - \frac{(l/R-\mu)^2}{\sigma^2} \right),
\end{equation}
with $b_0$ the height of the Gaussian, $\mu$ the mean value, and $\sigma$ the standard deviation. We display in Figure~\ref{fig:stat}(a) the histograms and their Gaussian fits corresponding to three values of the flow velocity. For $u_0 = 0$ the Gaussian mean value matches the actual $l/R$. As the flow velocity increases, the mean value is shifted towards larger $l/R$, but the width of the Gaussian is roughly the same regardless of the flow velocity.

\begin{figure}[!htp]
\centering
\includegraphics[width=.85\columnwidth]{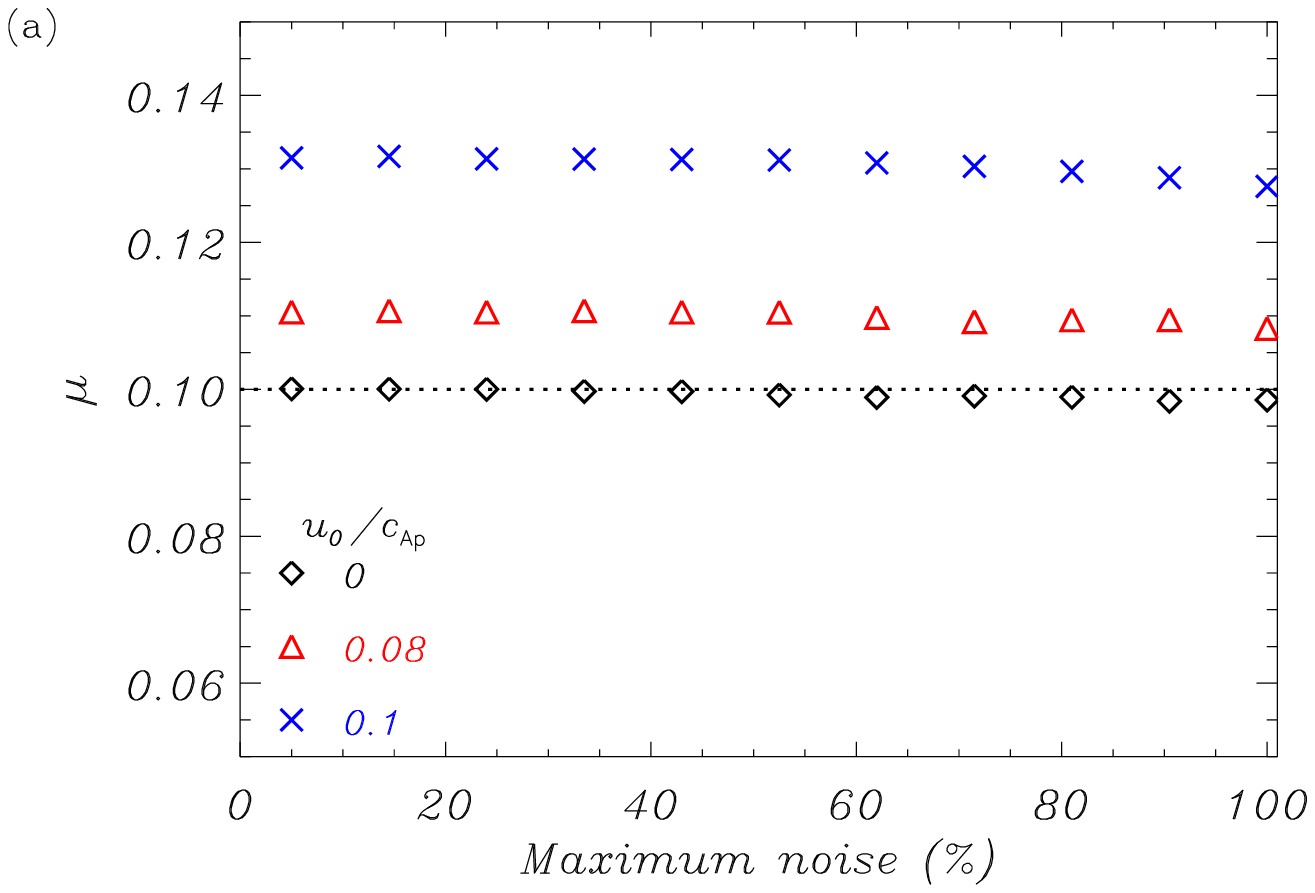}
\includegraphics[width=.85\columnwidth]{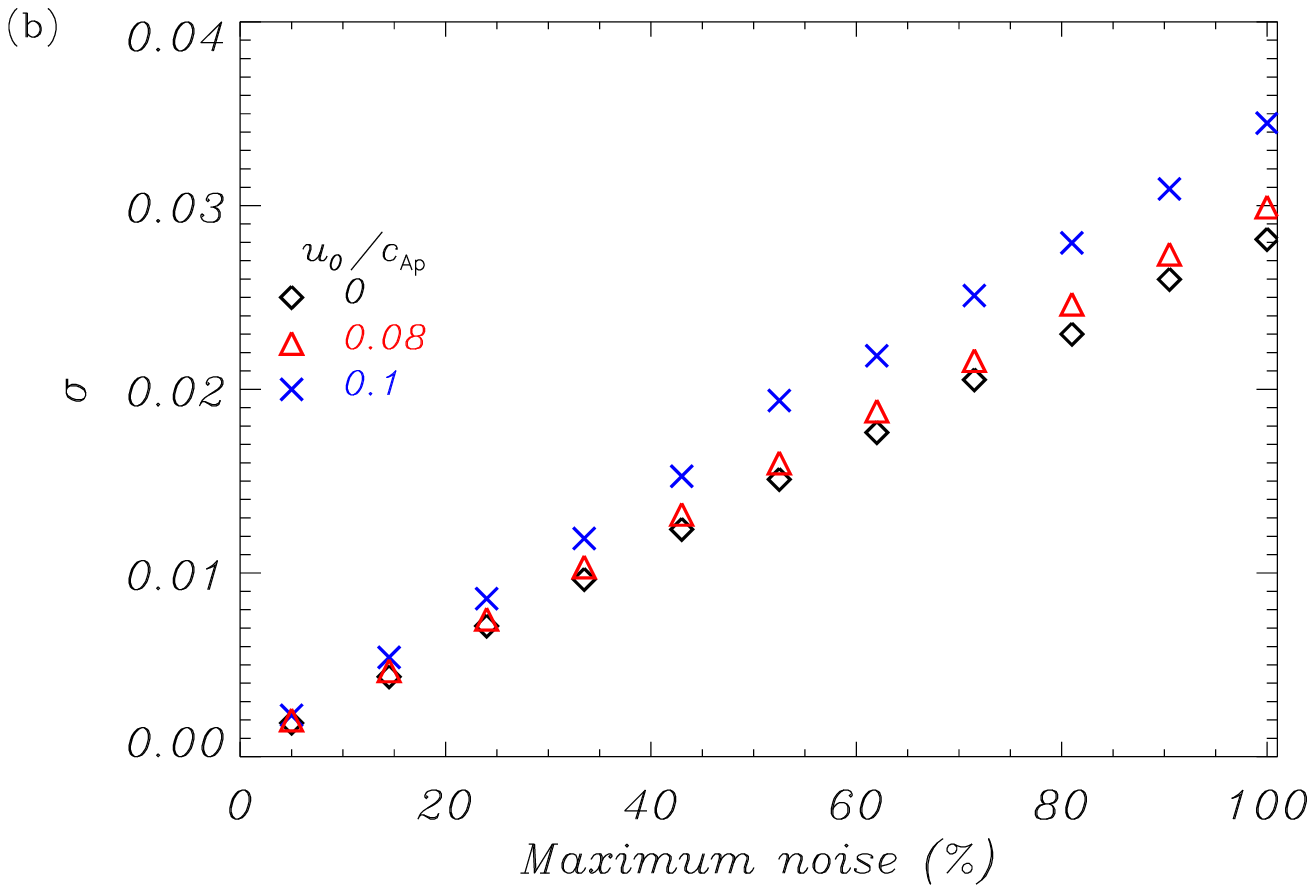}
\caption{(a) Mean value, $\mu$, and (b) standard deviation, $\sigma$, of the Gaussian distribution of seismologically estimated $l/R$ as functions of the percentage of background noise. The different symbols correspond to different flow velocities (indicated within the panels). The horizontal dotted line in panel (a) shows the actual value of $l/R$. \label{fig:stat2}}
\end{figure}

On the other hand,  Figure~\ref{fig:stat}(b) shows equivalent histograms but for a fixed value of the flow velocity and three different  $\left(\%\right)_{\rm noise}$. We find that the width of the Gaussian is strongly affected by noise, i.e., the larger the background noise, the wider the Gaussian. The width of the Gaussian is related to the uncertainty of $l/R$. Thus, we consistently obtain that the uncertainty increases when noise is increased, as expected. However, the mean value is not affected by the percentage of noise. The effect of noise is explored in more detail in Figure~\ref{fig:stat2}, which shows the fitted  $\mu$ (Figure~\ref{fig:stat2}a) and $\sigma$ (Figure~\ref{fig:stat2}b) as a function of the maximum background noise percentage for three flow velocities. We find that $\mu$ is independent of noise and its shift with respect to the actual $l/R$ is only determined by the flow velocity. For the set of parameters used in this test, the mean estimated $l/R$  when flow is present is larger than the actual value. On the contrary, the effect of flow on $\sigma$ is minor compared to the effect of noise. We obtain that $\sigma$ is a approximately linear with the maximum noise percentage. 

\begin{figure}[!htp]
\centering
\includegraphics[width=.85\columnwidth]{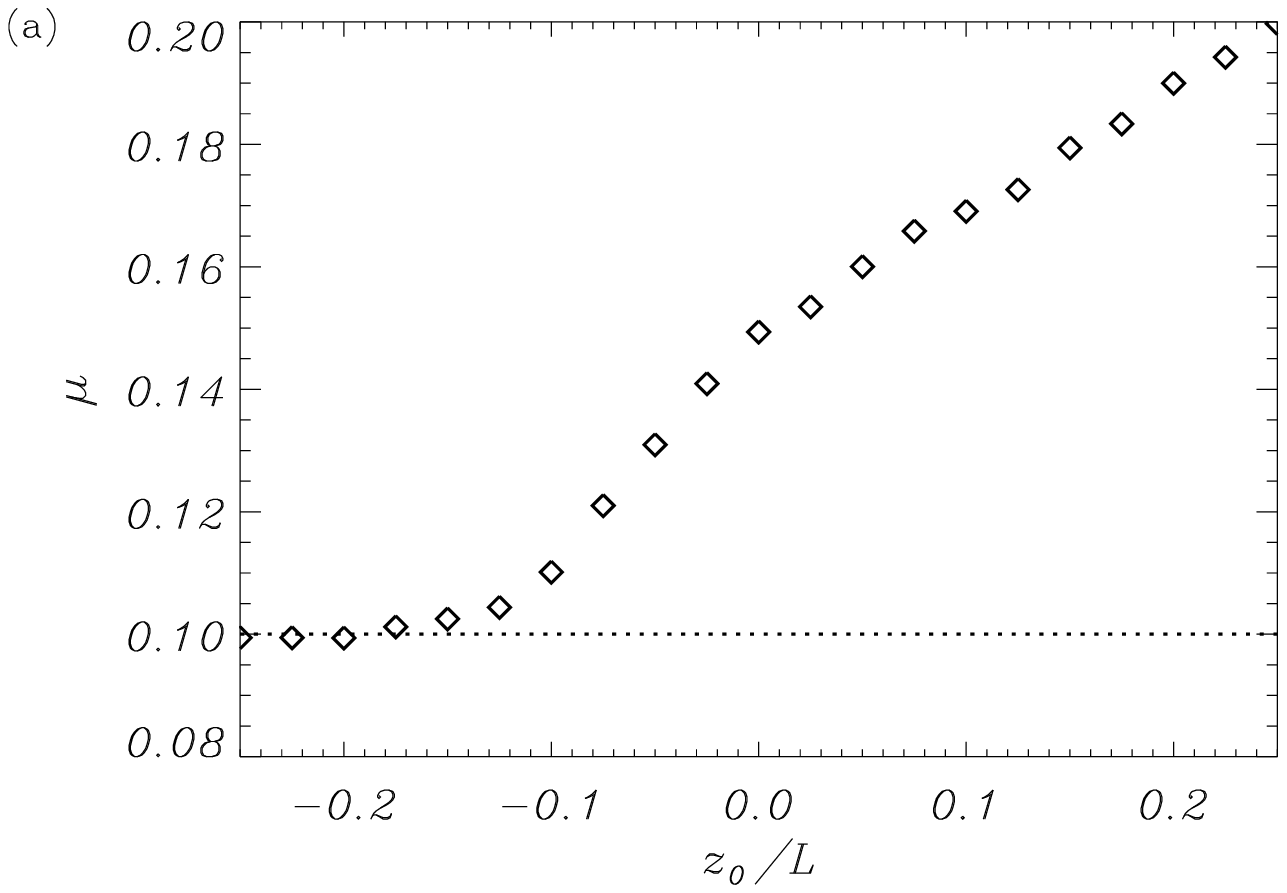}
\includegraphics[width=.85\columnwidth]{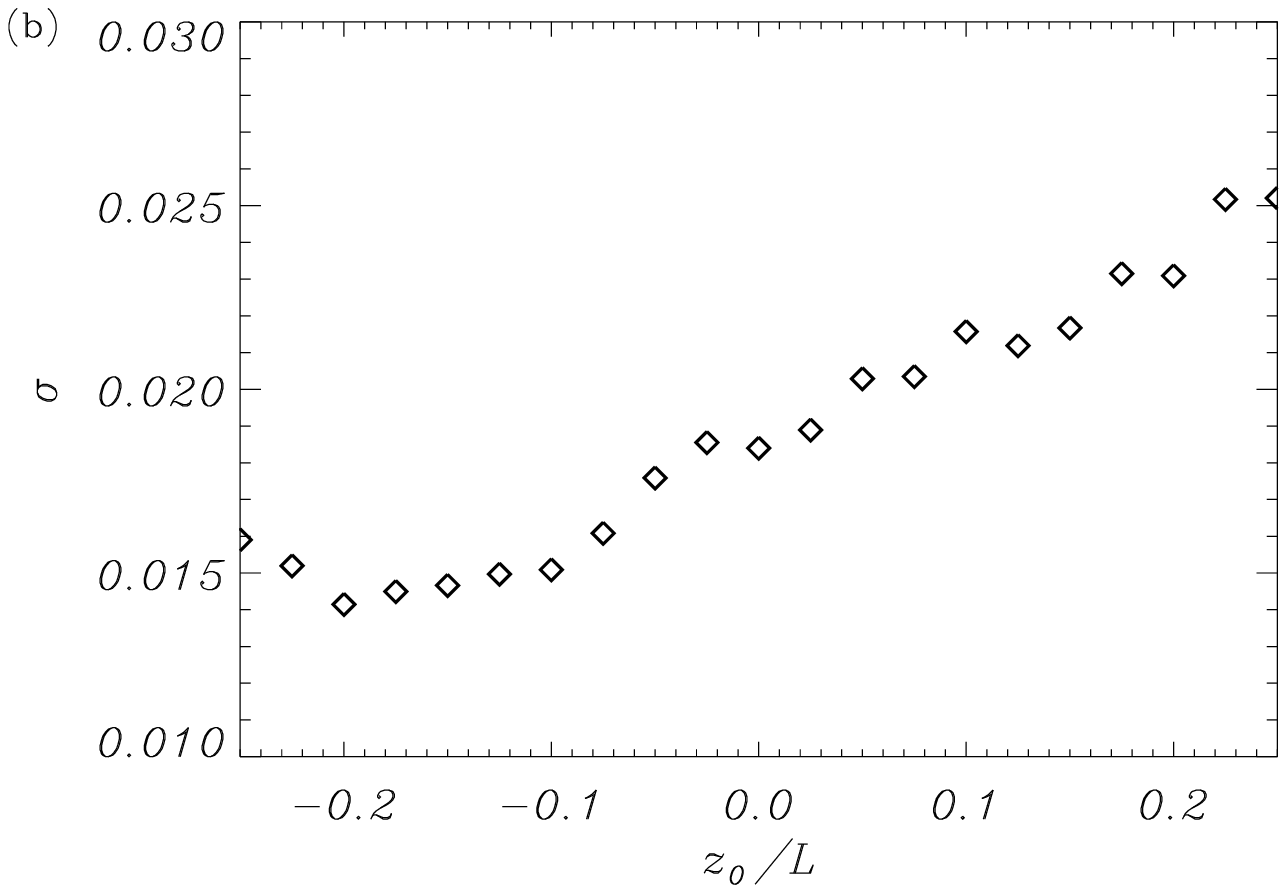}
\caption{(a) Mean value, $\mu$, and (b) standard deviation, $\sigma$, of the Gaussian distribution of seismologically estimated $l/R$ as functions of $z_0/L$. A maximum background noise of 50\% and $u_0 / \vap = 0.1$ are used. The horizontal dotted line in panel (a) shows the actual value of $l/R$. \label{fig:stat3}}
\end{figure}

Up to here we have obtained that the seismologically inferred $l/R$ tends to overestimate the actual value. Our goal now is to determine whether this is the general tendency or whether this is affected by the choice of the model parameters. There are several parameters that may affect the estimation of $l/R$. In particular, the role of $z_0$, i.e., the position of the thread with respect to the center of the magnetic tube  at the beginning of the oscillation, is worth being studied. To do so, we fix $u_0 / \vap = 0.1$, $\left(\%\right)_{\rm noise} = 50\%$, and select a particular value of $z_0 /L$  (the remaining parameters are the same as before). For a given $z_0 /L$ we repeat the previously explained procedure to generate the set of synthetic signals and the corresponding estimations of $l/R$. We fit the Gaussian function to the obtained histogram and compute $\mu$ and $\sigma$ as functions of $z_0 /L$. These results are displayed in Figure~\ref{fig:stat3}. Regarding $\mu$ (Figure~\ref{fig:stat3}a), we find that the estimated $l/R$ is larger than the actual value except when $z_0 /L \lesssim -0.1$. When $z_0$ is negative, i.e., the thread is initially located to the left of the center of the magnetic tube, the effect of flow is to increase the amplitude until the thread reaches the center of the tube, meaning that flow opposes resonant damping. This increase of the amplitude compensates the overestimation of $l/R$ by the fitting method, so that the mean estimated $l/R$ is closer to the actual value. On the contrary, when  $z_0 /L \gtrsim -0.1$ the flow contributes to damping during most of the evolution. This causes an overestimation of $l/R$. The behavior of $\sigma$ (Figure~\ref{fig:stat3}b) also shows an increasing trend with $z_0 /L$.

Based on the  results shown in this section, we conclude that flow and noise have two different effects on the estimation of $l/R$ using typical fitting methods in combination with the theoretical inversion formula. An exponentially damped harmonic function has been used as a proxy to the actual oscillation. On the one hand, flow tends to shift the estimated $l/R$ with respect to the actual value. The shift depends specifically on the flow velocity and on the position of the prominence thread at the beginning of the oscillation but, in general, an overestimation of $l/R$ is found in the statistical analysis. On the other hand, as expected, noise adds uncertainty to the estimation. This noise-related uncertainty might be partially reduced if some method able to subtract the background noise is applied to the raw data. However, some uncertainty intrinsically related to the effect of flow might remain even if the influence of noise is completely removed.

Real data are frequently affected by gaps in the time series. This issue has not been taken into account in the present analysis. The consideration of gaps would add more complexity to the analysis and, probably, would result in a new source of error. Appropriate methods accounting for the effect of gaps in the signals should be used \citep[see, e.g.,][]{carbonell1992}. This is worth being investigated in future works. 

\section{Conclusions}

\label{sec:conc}

In this paper we have studied the joint effect of resonant absorption and flow on the amplitude of standing kink waves in prominence threads.  The present work extends the previous paper by \citet{solerthreadflow} that did not take resonant damping into account. We have followed the method developed by \citet{ruderman2011a,ruderman2011b} to obtain a general analytic expression for the kink mode amplitude as a function of time which includes the effects of both resonant absorption and flow.

We find that flow and resonant absorption can either be competing effects on the amplitude or can both contribute to the damping of the kink mode depending on the instantaneous position of the dense thread within the prominence magnetic tube. For fast flows and short transverse inhomogeneity length scales the amplitude profile deviates from the classic exponential function for resonantly damped kink modes in static flux tubes. From the observational point of view, to determine the location of the dense plasma within the magnetic tube might be difficult since the  footpoints of the magnetic tube are not seen in the observations. 

The implications of our results for seismology of solar prominences have been explored. We have test the robustness of seismological estimates of the transverse inhomogeneity length scale. We have used synthetic data aiming to mimic real observations and have performed a statistical study. Our results show that the presence of flow can significantly affect the estimation of the transverse inhomogeneity length scale. Statistically, we find that this parameter is overstimated when an exponentially damped harmonic function, which does not take flow into account, is used to fit the actual oscillation. The presence of random background noise and/or intrumental errors adds further uncertainty to this estimation.

The seismology of flowing prominence threads using damped kink waves is more challenging than that of their static counterparts because of the effect of flow on the amplitude. The presence of flow adds complexity to the behavior of the oscillations and has a direct impact on seismology \citep[see, e.g.,][]{terradasflow}. Caution needs to be paid to the seismological estimates that do not take the influence of flow into account. 

\acknowledgements{
  Part of of this work was carried out when MSR was a guest in the Centre for Mathematical Plasma Astrophysics of KU Leuven. MSR acknowledges the warm hospitality of the Centre. RS thanks I. Arregui, J.L Ballester, and J. Terradas for reading the manuscript and for giving helpful comments. RS acknowledges support from a Marie Curie Intra-European Fellowship within the European Commission 7th Framework Program (PIEF-GA-2010-274716). RS and MG acknowledge the support from MICINN/MINECO and FEDER funds through grant AYA2011-22846.  MG acknowledges support from K.U. Leuven via GOA/2009-009. RS  acknowledges support from CAIB through the `grups competitius' scheme and FEDER Funds. MSR acknowledges financial support by the Leverhulme trust Senior Research Fellowship and by an STFC grant. Wavelet software was provided by C. Torrence and G. Compo, and is available at http://paos.colorado.edu/research/wavelets/}

\end{document}